\documentclass[preprint,preprintnumbers,amsmath,amssymb,superscriptaddress]{revtex4}

\include{ams}
\usepackage{amsmath,amssymb,amsthm}
\usepackage{dcolumn}
\usepackage{bm}
\usepackage{graphicx}
\usepackage{amsmath}
\usepackage{amssymb}
\newcommand{\sign}{\text{sign}}

\begin{document}

\title{Spin-dependent electron transmission across the corrugated graphene}

\author{M.Pudlak}
\affiliation{Institute of Experimental
Physics, Watsonova 47, 04001 Kosice, Slovakia}
\author{R.G.Nazmitdinov}
\affiliation{Bogoliubov Laboratory of
Theoretical Physics, Joint Institute for Nuclear Research, 141980
Dubna, Moscow region, Russia} \affiliation{Dubna State University,
141982 Dubna, Moscow region, Russia}
\begin{abstract}
We study various mechanisms of electron transmission across the
corrugations in the graphene sheet.
The spin dependence of the
electron transmission probability in the rippled graphene is
found.
The electrons mean free path and transmission
probabilities   are analysed for different distributions
of ripples in the graphene sheet as well.
We demonstrate that the periodically repeated
rippled graphene  structure (the superlattice) leads to the suppression
 of the
transmission of the ballistic electrons with one spin orientation in contrast to the other,
depending on the direction of the incoming electron flow.
\end{abstract}

\date{\today}

\maketitle

\section{Introduction}

Various experimental studies  indicate that a graphene sheet is
corrugated naturally due to intrinsic strains.
As a consequence,  one could, manipulating externally, alter the degree of
corrugation  to elucidate the  carrier transport response in
the graphene based system.
Nowadays, a few experimental techniques demonstrate
evidently a possibility of a spatial variation in graphene sheets.
As a result, it is found that ripples or wrinkles act as potential barriers for
charged carriers leading to their localization \cite{car16}.
In fact, it becomes possible
to form ripples without any change of doping by
means of the electrostatic manipulation  \cite{car19}.
Periodic nanoripples could be created with the aid
 of the chemical vapor deposition \cite{ni}.

The electronic structure of graphene based nanostructures  depends on their
geometry, indeed. In particular,  the effect of the
corrugations in graphene on the electronic structure and density of
states was evidently demonstrated  in Ref.\cite{Voz}.
The curvature of the graphene sheet affects
the $ \pi $ orbitals that determine the electronic properties.
Moreover, the surface curvature  enhances as well the effect of
the spin-orbit coupling. We recall that spin-orbit effects are commonly assumed to be
negligible in carbon based systems due to the weak atomic
spin-orbit splitting in carbon. However, as it was shown by Ando in the
effective-mass approximation \cite{Ando}, the spin-orbit becomes the essential property
in the carbon nanotube (CNT).
The modern understanding of transport through CNTs is reviewed in Ref.\cite{kouw}.

Evidently, a single ripple induces as well the spin-orbit coupling
due to its curvature, and, consequently, can be considered as a spin
scatterer. In fact, it is predicted that a ripple  could create in
graphene electron scattering, caused by the change in
nearest-neighbor hopping parameters by the curvature
\cite{Kat,Guinea}.  One-dimensional (1D) periodic rippled nanostructure
produces a strong focusing effect of ballistic electrons due the
Klein tunneling \cite{4}. Recently, it was found
that the spin-orbit interaction, raised by the surface curvature
\cite{Ando,PPN}, could produce a chiral transport \cite{PPN1}.
All these facts provide a solid basis for
the field of graphene transport properties, induced by variable
two-dimensional surface, itself. This field could be named as a
curvature induced electronics, i.e., {\it cuintronics}.

The purpose of the present paper is to study the spin
current manipulation in a rippled graphene sheet. We
consider the question as to how the electron mean free path  in graphene depends on the
electron spin, assuming a random distribution of the ripple positions. Further,
the transport of ballistic electrons through a graphene based 1D superlattice will be analysed,
when the superlattice consists of ripple+flat pieces that are periodically repeated.

\section{Hamiltonian}
\begin{figure}
\centerline{\includegraphics{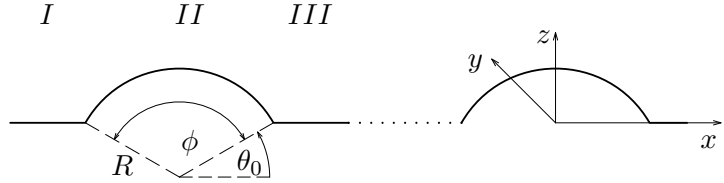}\unitlength=1mm\begin{picture}(0,0)(0,0)
\put(-55,20.5){\makebox(0,0)[b]{$III$}}
\put(-71,20.5){\makebox(0,0)[b]{$II$}}
\put(-90,20.5){\makebox(0,0)[b]{$I$}}
\put(-63,0.5){\makebox(0,0)[b]{$\theta_0$}}
\put(-71,3.){\makebox(0,0)[b]{$\phi$}}
\put(-80,0.5){\makebox(0,0)[b]{$R$}}
\put(-2,4){\makebox(0,0)[b]{$x$}}
\put(-33,14){\makebox(0,0)[b]{$y$}}
\put(-24,18){\makebox(0,0)[b]{$z$}}
\end{picture}}
\caption{The corrugated graphene system. The flat region (between
two arcs) has the length $L_{1}$, while the arc length is  $L_{2}$.
The ripples are ordered in $x$ direction with the symmetry $y$-axis
.} \label{medium}
\end{figure}

The corrugated graphene structure is modelled by
a curved surface in a form of arc of a circle connected from
the left-hand and the right-hand sides to two flat graphene sheets (see Fig.\ref{medium}).
The solution for a flat graphene is well
known \cite{1,2}. The solution for a curved graphene surface can be
expressed in terms of the results obtained for  the armchair CNT
in an effective mass approximation, when only the interaction
between nearest neighbour atoms is taken into account \cite{PPN}.
 The axis $y$ is chosen as the symmetry axis .

We briefly recapitulate the major results \cite{PPN} in the vicinity of the Fermi level
$E=0$ for the point $K$ in the
presence of the curvature induced spin-orbit interaction in the armchair CNT.
In this case the eigenvalue problem is defined as \cite{Ando}

\begin{equation}
\label{2a} \left(\begin{array}{cc}0&\gamma
(\hat{k}_{x}-i\hat{k}_{y})+i\frac{\delta
\gamma^{'}}{4R}\sigma_{x}(\vec{r})-\frac{2\delta \gamma
p}{R}\sigma_{y}
\\\gamma (\hat{k}_{x}+i\hat{k}_{y})-i\frac{\delta
\gamma^{'}}{4R}\sigma_{x}(\vec{r})-\frac{2\delta \gamma
p}{R}\sigma_{y}&0\end{array}\right)\left(\begin{array}{c}
F^{K}_{A}\\F^{K}_{B}\end{array}\right)=E
\left(\begin{array}{c}F^{K}_{A}\\F^{K}_{B}\end{array}\right)
\end{equation}

with $\hat{k}_{x}=-i\frac{\partial}{R\partial
\theta}$,$\hat{k}_{y}=-i\frac{\partial}{\partial y}$,
$\sigma_{x}(\vec{r})=\sigma_{x}\cos\theta -\sigma_{z}\sin\theta$,
and
\begin{equation}
F^{K}_{A}=\left(\begin{array}{c}F^{K}_{A,\uparrow}\\F^{K}_{A,\downarrow}\end{array}\right),
F^{K}_{B}=\left(\begin{array}{c}F^{K}_{B,\uparrow}\\F^{K}_{B,\downarrow}\end{array}\right)
\end{equation}
The parameter $\gamma
=-\sqrt{3}V_{pp}^{\pi}a/2=\sqrt{3}\gamma_{0}a/2$,
$\gamma^{'}=\sqrt{3}(V_{pp}^{\sigma}-V_{pp}^{\pi})a/2$, where
$V_{pp}^{\pi}$ and $V_{pp}^{\sigma}$ are the transfer integrals for
$\pi$ and $\sigma$ orbitals, $a$ is the length of the primitive
translation vector ( $a=\sqrt{3}d\simeq 2.46$\AA{}, where $d$ is
the distance between atoms in the unit cell). Here, the following notations
have been introduced:
\begin{eqnarray}
&p=1-3\gamma^{'}/8\gamma\,, \quad
\delta = \Delta/3\epsilon_{\pi\sigma}\,, \\
&\epsilon_{\pi\sigma}=\epsilon_{2p}^{\pi}-\epsilon_{2p}^\sigma\,,\quad
\Delta = i\frac{3\hbar}{4m^{2}c^{2}}\langle x|\frac{\partial
V}{\partial x} p_{y}-\frac{\partial V}{\partial y} p_{x}|y\rangle\,.
\end{eqnarray}
The energy $\epsilon_{2p}^{\sigma}$ is the energy of
$\sigma$-orbitals which are localized between carbon atoms. The
energy $\epsilon_{2p}^{\pi}$ is the energy of $\pi$-orbitals which
are directed perpendicular to the nanotube surface. It is assumed
that $\gamma_{0}=-V_{pp}^{\pi}\approx 3$ eV and $V_{pp}^{\sigma}\approx 5$ eV. The
eigenvalues and eigenfunctions of Eq.(\ref{2a})
 are discussed in detail in \cite{PPN}. It is assumed that the parameter $\delta$ is
defined in the range $(10^{-3},10^{-2})$, and $|p|\leq 0.1$ \cite{Ando}.
In the numerical computation we use the following values: $\delta =0.01$ and $p=0.1$.
The solutions of the Eq.(\ref{2a}) are employed to describe the electron state in the ripple region.

\section{Transmission through one ripple}

To analyse the scattering problem  through the corrugated
graphene structure
we consider three regions.
The flat graphene pieces are located
in: the region (I), $-\infty<x< -R\cos\theta_{0}$,  $-\infty < y < \infty$;
and the region (III), $R\cos\theta_{0}<x<\infty$, $-\infty < y < \infty$  (see Fig.\ref{medium}).
 In the region (II), $-R\cos\theta_{0} <x< R\cos\theta_{0}$ and $-\infty < y < \infty$, we consider
a ripple (a curved surface in a form of arc of a circle).
Hereafter, we consider a wide enough graphene sheet W$\gg$M, where W and M being, respectively,
as the width along the y axis and the length along x axis of the graphene sheet.
It means that we keep the translational invariance along the y axis and neglect the edge effects.
Here, we assume that ballistic electrons are injected to the ripple in a
perpendicular direction, i.e., $k_{y}=0$.

The eigenenergy and eigenfunction in the flat
graphene pieces are defined as
\begin{equation}
\label{5} E=\pm \gamma |k|\,,
\end{equation}
\begin{equation}
\label{23} \Psi_{0}^{\sigma}(x,k)=\frac{1}{2}\left(\begin{array}{c}
\sign(k E) \\1
\end{array} \right) \otimes \left(\begin{array}{c} 1
\\ \sigma i\end{array}\right)e^{i k x},\quad \sigma =\pm\,.
\end{equation}
The eigenfunctions and eigenenergies in the region II can be expressed in the form \cite{PPN1}
\begin{equation}
E_{s}=\pm \sqrt{t_{m}^{2}+\lambda_{y}^{2}}+ s \lambda_{x}, \quad s=\pm\,,
\end{equation}
\begin{equation}
\label{25d}
\Psi_{A}^{s}(\theta,m)=\frac{1}{\sqrt{2(A_{s}^{2}+1)}}\left(\begin{array}{c}
\Phi_{A}^{s}(\theta)
\\ -s \sigma_{y}\Phi_{A}^{s}(\theta)
\end{array}\right)e^{i m  \theta}\,,
\end{equation}
\begin{equation}
\label{26d} \Phi_{A}^{s}(\theta)=e^{-i\frac{\theta}{2}
\sigma_{y}}\left(\begin{array}{c} A_{s}
\\ -s i\end{array}\right), \quad A_{s}=\left(s \lambda_{y}
+\sqrt{t_{m}^{2}+\lambda_{y}^{2}}\right)/t_{m}\,,
\end{equation}
where
\begin{equation}
\label{def}
t_{m}= \frac{\gamma}{R}m\,,\quad \lambda_{y}
=\frac{\delta\gamma^{'}}{4R}\,,\quad \lambda_{x}
=\frac{\gamma}{R}(\frac{1}{2}+2\delta p)\,.
\end{equation}
In fact, the parameters $\lambda_{x,y}$ determine the strength of the
spin-orbit coupling induced by the curvature \cite{Ando,PPN}.
In our consideration
the eigenstates of the Pauli matrix $\sigma_{y}$ are associated with
the eigenfunctions with the spin up and down.
These solutions are used to describe the transmission and the
reflection of electrons on the ripple.

We suppose that the incident electron moves from the left planar part  with
spin up ($\sigma =+$) to the right planar part  along the x axis,
and its energy is the integral of motion.
 The unknown reflection and transmission amplitudes $r_{\alpha}^{\beta},
t_{\alpha}^{\beta}$($\alpha,\beta=\uparrow, \downarrow$) are determined by requiring
the continuity of the wave function $\Psi$ at the boundaries between the flat and corrugated
graphene regions. Consequently, we have the following condition at the boundary between region I and II
\begin{eqnarray}
\label{27}
&&\Psi_{0}^{+}(x,k)+r(L)^{\uparrow}_{\uparrow}\Psi_{0}^{+}(x,-k)+r(L)^{\uparrow}_{\downarrow}\Psi_{0}^{-}(x,-k)=
a_{+}\Psi_{A}^{+}(\theta,m_{+})+b_{+}\Psi_{A}^{+}(\theta,-m_{+})\nonumber\\
&&+a_{-}\Psi_{A}^{-}(\theta,m_{-})+
b_{-}\Psi_{A}^{-}(\theta,-m_{-}) ;\,\quad x=-R\cos\theta_{0}\,, \quad \theta =\pi+\theta_{0}\,.
\end{eqnarray}
The quantum numbers $m_{+}$ and $m_{-}$ are determined by the equation
\begin{equation}
E=\gamma k =\pm\sqrt{\left(\gamma m_{+}/R\right)^{2}+\lambda_{y}^{2}}+
\lambda_{x}=\pm\sqrt{\left(\gamma
m_{-}/R\right)^{2}+\lambda_{y}^{2}}- \lambda_{x}\,,
\end{equation}
where $E$ is the incoming electron energy. As a result, we have
\begin{equation}
\label{12} m_{s} =\pm
\frac{R}{\gamma}\sqrt{(E-s\lambda_{x})^{2}-\lambda_{y}^{2}}\,,\quad s=\pm\,.
\end{equation}
At the boundary between region II and III we have the condition
\begin{eqnarray}
&a_{+}\Psi_{A}^{+}(\theta,m_{+})+b_{+}\Psi_{A}^{+}(\theta,-m_{+})+a_{-}\Psi_{A}^{-}(\theta,m_{-})+
b_{-}\Psi_{A}^{-}(\theta,-m_{-})=\nonumber\\
\label{28}
&t(L)^{\uparrow}_{\uparrow}\Psi_{0}^{+}(x,k)+
t(L)^{\uparrow}_{\downarrow}\Psi_{0}^{-}(x,k);\, \quad x=R\cos\theta_{0}\,,\quad \theta =2\pi-\theta_{0}\,.
\end{eqnarray}
Solving Eqs.(\ref{27},\ref{28}), we obtain the following result for the transmission
probability with the conserved spin polarization
\begin{equation}
|t(L)_{\uparrow}^{\uparrow}|^{2}=\frac{1}{z_{-}^{2}(\phi)+1}\,,
\end{equation}
where $\phi =\pi-2\theta_{0}$ and
\begin{equation}
z_{\pm}(\phi)=\frac{\lambda_{y}\sin(m_{\pm}\phi)}{t_{m_{\pm}}}\,.
\end{equation}
The probability  of the reflected electron (by the ripple)  with
the spin-flip is defined as
\begin{equation}
|r(L)_{\downarrow}^{\uparrow}|^{2}=\frac{z_{-}^{2}(\phi)}{z_{-}^{2}(\phi)+1}\,.
\end{equation}
The probabilities of the reflection without the spin-flip and
the transmission with the spin-flip are
zero, i.e., $|r(L)_{\uparrow}^{\uparrow}|^2=|t(L)_{\downarrow}^{\uparrow}|^2=0$.
Thus, the corrugated structure yields the backscattering
with the spin-flip phenomenon.

Let us consider the case when the
incoming electron has the spin down. The wave function of the incoming
electron (in the flat graphene) has the form
\begin{equation}
\label{28a} \Psi_{0}^{-}(x,k)=\frac{1}{2}\left(\begin{array}{c}
1 \\
1
\end{array}\right)\otimes \left(\begin{array}{c} 1 \\ -i
\end{array}\right)e^{ikx}\,.
\end{equation}
Following the same procedure as above, we obtain
 that the electron
is transmitted through the ripple without the spin-flip with the
probability
\begin{equation}
|t(L)_{\downarrow}^{\downarrow}|^{2}=\frac{1}{z_{+}^{2}(\phi)+1}\,,
\end{equation}
while the reflection with the spin-flip is defined by the probability
\begin{equation}
|r(L)_{\uparrow}^{\downarrow}|^{2}=\frac{z_{+}^{2}(\phi)}{z_{+}^{2}(\phi)+1}\,.
\end{equation}
Here, as in the previous case, we have
$|r(L)_{\downarrow}^{\downarrow}|^2=|t(L)_{\uparrow}^{\downarrow}|^2=0$.
For the electron, coming from
the right side, we have the following results:
\begin{eqnarray}
&|t(R)_{\downarrow}^{\downarrow}|^{2}=|t(L)_{\uparrow}^{\uparrow}|^{2}\,,\quad
|r(R)_{\uparrow}^{\downarrow}|^{2}=|r(L)_{\downarrow}^{\uparrow}|^{2}\,,\\
&|t(R)_{\uparrow}^{\uparrow}|^{2}=|t(L)_{\downarrow}^{\downarrow}|^{2}\,, \quad
|r(R)_{\downarrow}^{\uparrow}|^{2}=|r(L)_{\uparrow}^{\downarrow}|^{2}\,.
\end{eqnarray}
Thus, the direction exchange of the incoming electron flow to the corrugated graphene structure
yields the change of the dominance  of one or another spin polarization of
the transmitted current.
In other words, we observe that the corrugated graphene structure supports the chiral symmetry
for the transmitted electron flow.

\section{incoherent scattering}

The above results imply that, if there is some concentration of ripples in the  nanostructure,
 the mean free path for electrons with different spin orientations
may be different. To elucidate this point, we consider
a sample of the length $L_{0}$, containing an average concentration
of ripples ($n$). It is natural to assume  that ripples are distributed randomly.
For the sake of discussion, we neglect the interference between successive scatterers.
We return to this point later.

The right/left directed
$I_{+}(x)/I_{-}(x)$ electron flows (that are position dependent)
at the distance $x$ are subject to the equations
\begin{equation}
\frac{dI_{+}(x)}{dx}=-\xi I_{+}(x)+\lambda I_{-}(x)\,.
\end{equation}
Here,
\begin{equation}
\xi =\mid r^{\uparrow}_{\downarrow}(L)\mid^{2}n\,,\quad
\lambda=\mid r^{\downarrow}_{^{\uparrow}}(R)\mid^{2}n\,,
\end{equation}
where $n dx$ is the number of ripples between $x$ and $x+dx$.
The condition
$\mid r^{\uparrow}_{\downarrow}(L)\mid^{2}=\mid r^{\downarrow}_{^{\uparrow}}(R)\mid^{2}$
yields the equivalence $\xi=\lambda$.
Consequently, the mean free path (m.f.p.) for the  backscattering is
$1/\xi$. As a result, we have for the m.f.p.
 for electrons (coming from the left to the right side):
 \begin{itemize}
 \item
$l^{\downarrow}=1/{\mid r(L)_{\uparrow}^{\downarrow}\mid^{2}n}$ (with the spin down)\,;

\item
$l^{\uparrow}=1/{\mid r(L)_{\downarrow}^{\uparrow}\mid^{2}n}$ (with the spin up)\,,
\end{itemize}
where the backscattering coefficients $r^{\alpha}_{\beta}(L)$
depend on the energy of the incoming electron.

For the steady states, when
\begin{equation}
I_{+}(x)-I_{-}(x)=I \to const.\,,
\end{equation}
we
obtain for the right directed flow (see also Ref.\cite{Lund})
\begin{equation}
I_{+}(x=L_{0})=T^{\alpha}I_{+}(x=0)\,.
\end{equation}
Here, the transmission probability $T^{\alpha}$ has the form
\begin{equation}
T^{\alpha}=\frac{l^{\alpha}}{l^{\alpha}+L_{0}}=\frac{1}{1+\mid
r(L)_{\beta}^{\alpha}\mid^{2}n L_{0}}\,; \quad \alpha, \beta
=\uparrow\,,\downarrow\,.
\end{equation}
In the case of the electron flux moving from the left to the right planar piece
through the ripple,
numerical results demonstrate the dominance of the transmission of the electron
flow with spin up electrons in a particular energy window (see Fig.\ref{inc}).
Thus, the electron transmission across the corrugation
region depends on the electron spin polarization.

\begin{figure}
\label{inc}
\includegraphics[height=6cm,clip=]{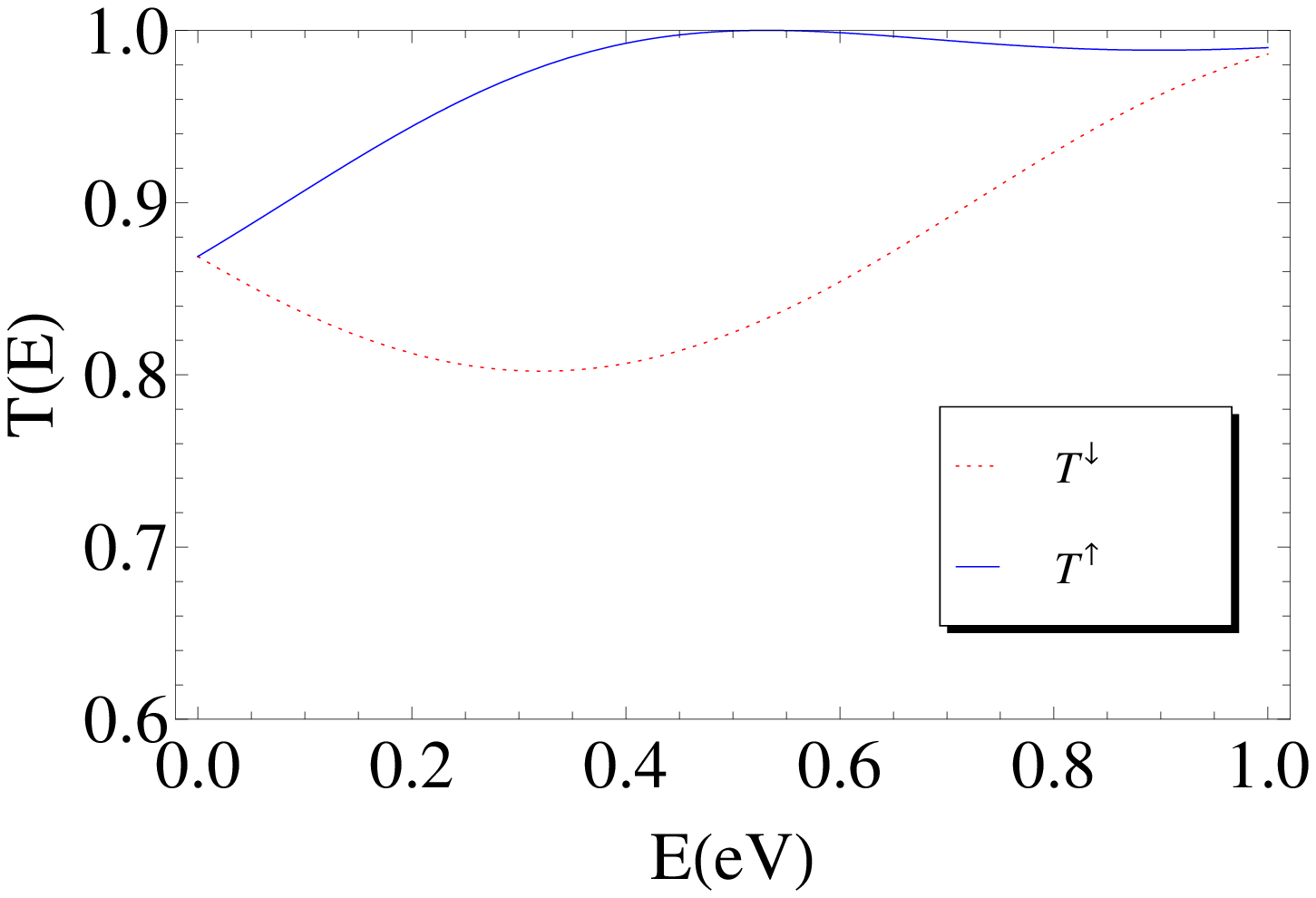}
\begin{picture}(0,0)(0,0)
\put(-55,130){\makebox(0,0)[b]{$(a)$}}
\end{picture}\
\includegraphics[height=6cm,clip=]{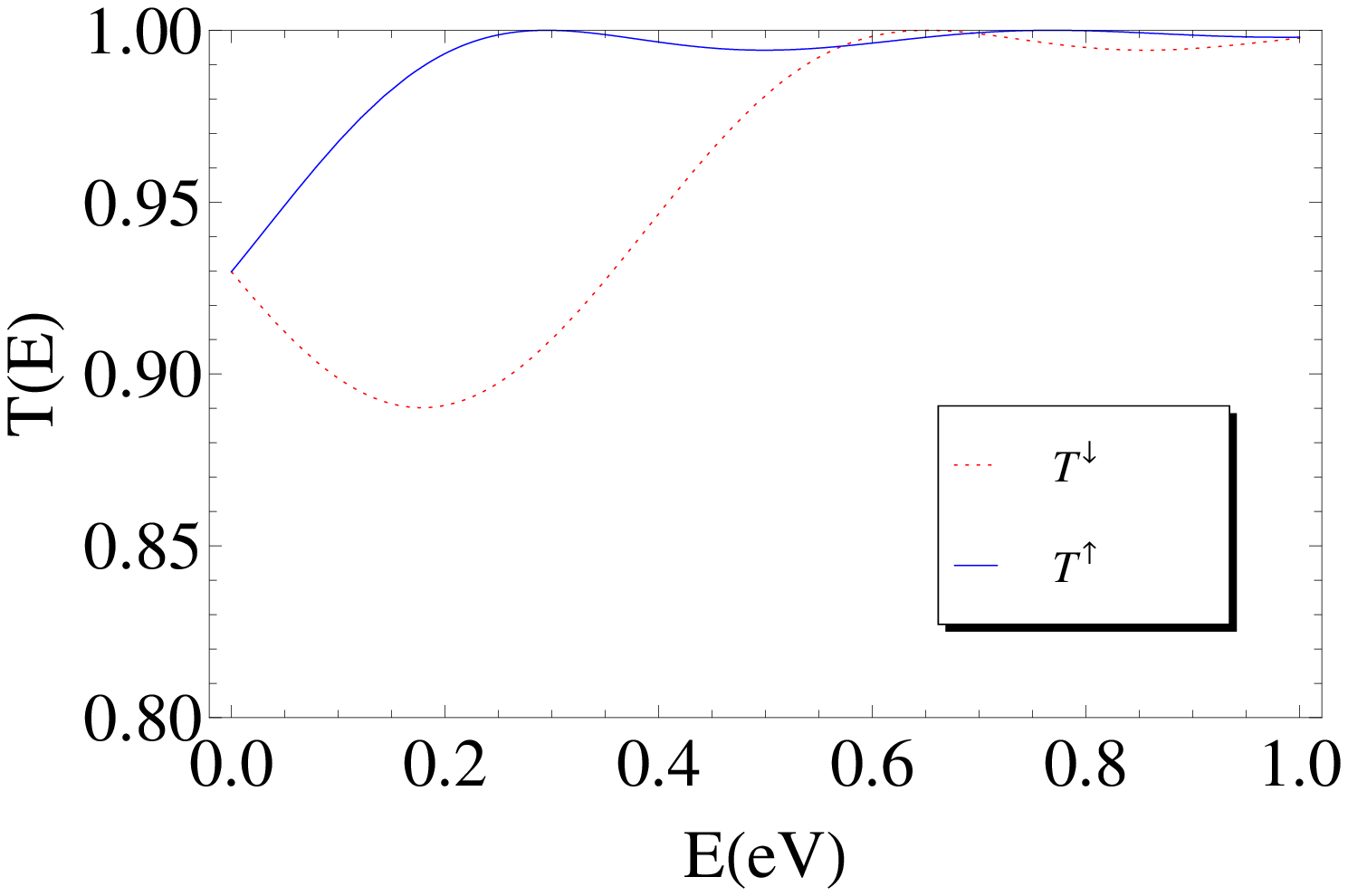}
\begin{picture}(0,0)(0,0)
\put(-55,130){\makebox(0,0)[b]{$(b)$}}
\end{picture}
\caption{(Color online) Incoherent electron scattering.
The electron flux moves from the left to the right planar piece through
the ripple.
Transmission probability $T$ as a
function of the incoming electron energy :
for electron with spin down (dotted line) and spin
up (solid line). The sample
length is $L_{0}=5\mu m$; the angle $\phi =3\pi/4$;  (a) the ripple radius
$R=10$\AA {},  the ripple density $n=200 \mu m^{-1}$; (b) the ripple radius
$R=18$\AA {}, the ripple density $n=100 \mu m^{-1}$. }
 \label{inc}
 \end{figure}

We recall that the ripples are an inherent feature of graphene.
It appears that the electron reflection caused by the spin-orbital coupling, induced
by the ripple curvature, is the important factor. Indeed, it determines spin dependent mean free path
in the graphene sheet, which is naturally corrugated.

\section{Transmission across N ripples}
\subsection{Analytical results}

The ability of the modern technology to
create periodic rippled graphene structures (mentioned in Introduction)
raises the question as to how the electron transport would be affected
by N periodically distributed ripples. In this section we  discuss the electron
scattering in the graphene based superlattice.
Each element of the periodic chain consists of the the ripple connected
to the flat graphene piece (see Fig.\ref{medium}).
The element includes the regions II and III.
There are various approaches that allow to treat analytically this
problem (e.g. \cite{Gom}).
Our approach is based on the transfer matrix method, similar in spirit
to the one discussed in Ref.\cite{cvet}.
And below we follow it to obtain our analytical results.

Let us consider the electron beam with the spin up, which is moving from the left to
the right side of the superlattice.
Using the continuity conditions on the boundaries, we obtain the equations for the
transmission and corresponding reflection probabilities through the
block of $N$ ripples in the form
\begin{equation}
\label{51}
\left(\begin{array}{c}1\\
r(L)^{\uparrow}_{\downarrow}
\end{array}\right)=\left(\begin{array}{cc}A_{11}&A_{12}\\A_{21}&A_{22}
\end{array}\right)^{N}\left(\begin{array}{c}
t(L)^{\uparrow}_{\uparrow}\\0\end{array}\right)\,,
\end{equation}
\begin{equation}
A_{11}=e^{-i\Omega_{-}}\left(\cos m_{-}\phi-i
\frac{A_{-}^{2}+1}{2A_{-}}\sin  m_{-}\phi\right)=A_{22}^{*}\,,
\end{equation}
\begin{equation}
A_{12}=-i e^{-i\Omega_{-}}\frac{A_{-}^{2}-1}{2A_{-}}\sin m_{-}\phi
=A_{21}^{*}\,,
\end{equation}
where $\Omega_{-}=kL_{1} - \phi /2$.
Taking into account that there is the unitary transformation $U$ which diagonalizes the
matrix A, i.e.,
\begin{equation}
U^{-1}AU=\left(\begin{array}{cc}\lambda_1&0\\0&\lambda_{2}
\end{array}\right)\,,
\end{equation}
we can introduce the following notations
\begin{equation}
\label{52} \left(\begin{array}{cc}A_{11}&A_{12}\\A_{21}&A_{22}
\end{array}\right)^{N}=\left(\begin{array}{cc}N_{11}&N_{12}\\N_{21}&N_{22}
\end{array}\right)
\end{equation}
with the following elements:
\begin{equation}
N_{11}=\frac{A_{11}[(\lambda_{1}^{\sigma})^{N}-(\lambda_{2}^{\sigma})^{N}]+
(\lambda_{2}^{\sigma})^{N-1}-(\lambda_{1}^{\sigma})^{N-1}}{\lambda_{1}^{\sigma}-\lambda_{2}^{\sigma}}
=N_{22}^{*}\,,
\end{equation}
\begin{equation}
\label{n12}
N_{12}=A_{12}A_{N}^{\sigma}=N_{21}^{*}\,,\quad
A_{N}^{\sigma}=\frac{(\lambda_{2}^{\sigma})^{N}-
(\lambda_{1}^{\sigma})^{N}}{\lambda_{2}^{\sigma}-\lambda_{1}^{\sigma}}\,,
\quad \sigma=-\,.
\end{equation}
This trick determines the eigenvalues
\begin{eqnarray}
\label{l12}
&&\lambda_{1,2}^{(-)}=c\pm\sqrt{c^2-1}\,,\quad c=(A_{11}+A_{22})/2=\\
\label{23}
&&
=\cos\Omega_{-}\cos
m_{-}\phi-\frac{A_{-}^{2}+1}{2A_{-}}\sin m_{-}\phi \sin\Omega_{-}\,.
\end{eqnarray}
Using the result $|N_{11}|^{2}-|N_{12}|^{2}=1$, we obtain
\begin{equation}
\label{cup}
T^{\uparrow}=|t(L)^{\uparrow}_{\uparrow}|^{2}=\frac{1}{1+[z_\sigma (\phi)A_{N}^{\sigma}]^2}\,,\quad \sigma=-\,.
\end{equation}

Following the same steps [see Eqs.(\ref{51})-(\ref{n12})] for the electrons with the spin down, we obtain
\begin{equation}
\label{cdown}
T^{\downarrow}=|t(L)^{\downarrow}_{\downarrow}|^{2}=
\frac{1}{1+[z_\sigma (\phi)A_{N}^{\sigma}]^2}\,,\quad \sigma=+\,.
\end{equation}
where
\begin{eqnarray}
&\lambda_{1,2}^{(+)}=c \pm \sqrt{c^{2}-1}\,,\\
\label{23}
&c=\cos\Omega_{+}\cos
m_{+}\phi-\frac{A_{+}^{2}+1}{2A_{+}}\sin m_{+}\phi \sin\Omega_{+}\,,
\end{eqnarray}
and $\Omega_{+}=kL_{1} + \phi /2$. We also have
$|r(L)^{\uparrow}_{\downarrow}|^{2}=1-|t(L)^{\uparrow}_{\uparrow}|^{2}$,
$|r(L)^{\downarrow}_{\uparrow}|^{2}=1-|t(L)^{\downarrow}_{\downarrow}|^{2}$.
The reflection probability without the spin-flip $|r(L)_{\alpha}^{\alpha}|^{2}=0$,
as well as the transmission probability with the spin-flip
$|t(L)_{\alpha}^{\beta}|^{2}=0$.

For the electron flow, moving from the right to the left side of the superlattice,
we have the following relations:
\begin{eqnarray}
&&|t(R)_{\alpha}^{\alpha}|^{2}=|t(L)_{\beta}^{\beta}|^{2}\,,\\
&&|r(R)_{\alpha}^{\beta}|^{2}=|r(L)_{\beta}^{\alpha}|^{2}\,.
\end{eqnarray}
 All these
results are fulfilled for the conducting band $(E>0)$. For the valence
electrons $(E<0)$, the channels, through which the electron transfer
occurs, are inversed with respect to those of the conductance band.
In particular, the transmission probability $|t(L)^{\uparrow}_{\uparrow}|^{2}$
is determined by Eq.(\ref{cdown}), while the transmission probability
$|t(L)^{\downarrow}_{\downarrow}|^{2}$ is determined by Eq.(\ref{cup})
for $E<0$.

\subsection{Numerical results}

From the above analysis of the single ripple it follows, that  in some
energy interval the transport for the ballistic electrons with one spin polarization is more transparent
than for those with the opposite spin polarization.
To gain a better insight into the effect of the superlattice on the electron
transport, we investigate numerically its dependence on:
i) the number N of the ripples; ii)the radius of the ripple; iii)the angle of the ripple;
iii)the distance between ripples.

\begin{figure}
\includegraphics[height=6cm,clip=]{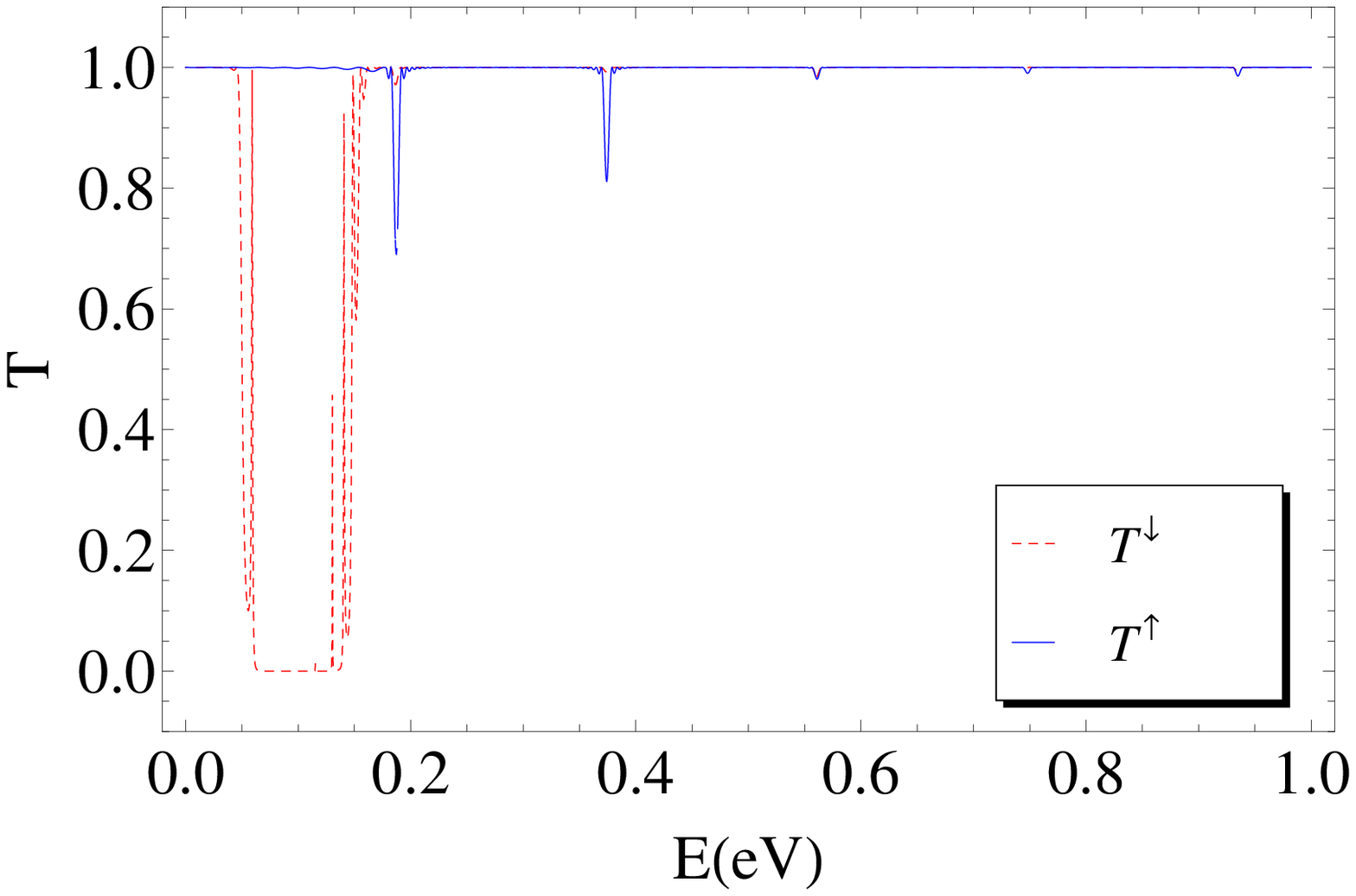}
\begin{picture}(0,0)(0,0)
\put(-65,130){\makebox(0,0)[b]{$(a)$}}
\end{picture}\
\includegraphics[height=6cm,clip=]{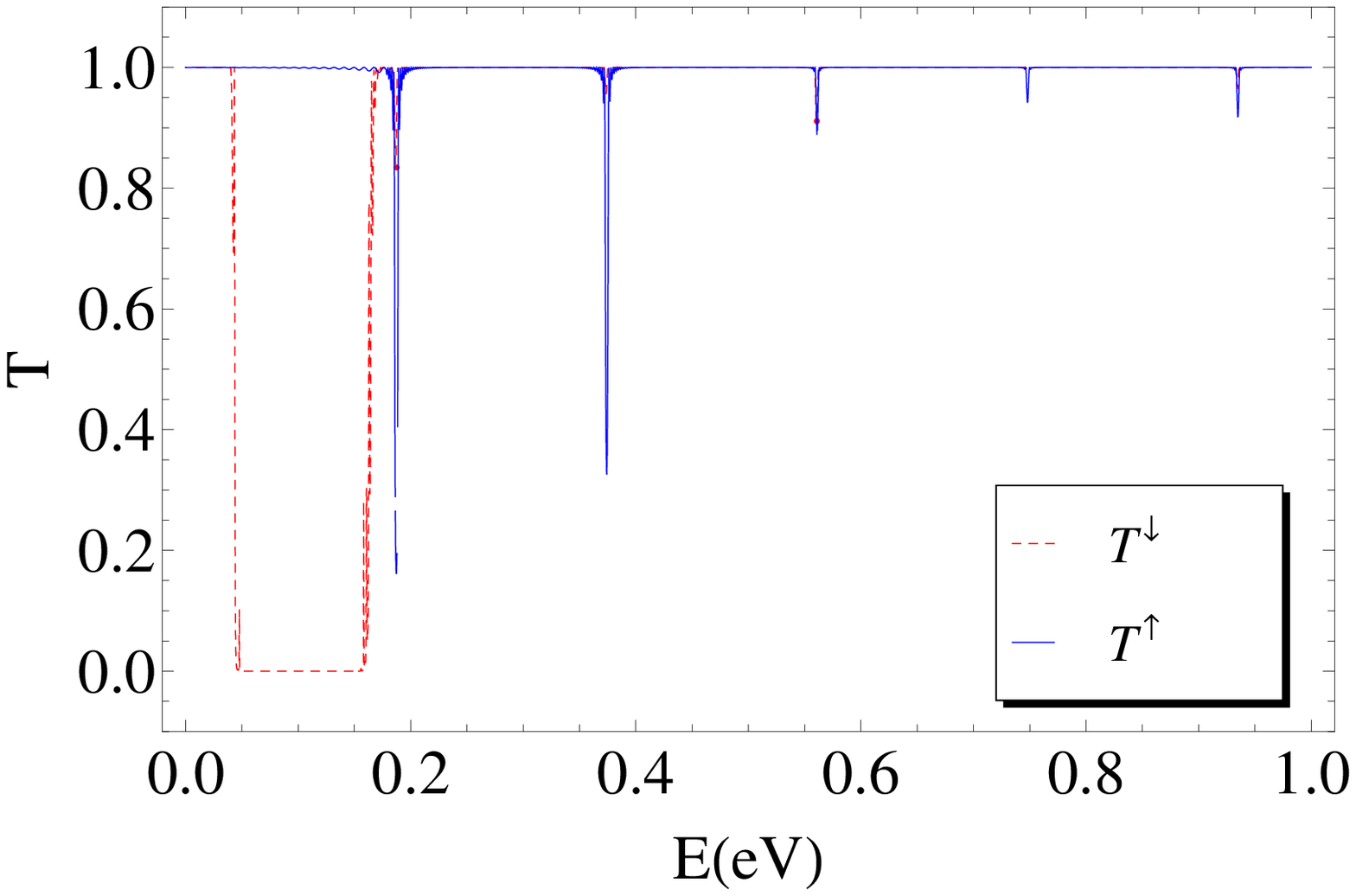}
\begin{picture}(0,0)(0,0)
\put(-45,130){\makebox(0,0)[b]{$(b)$}}
\end{picture}
\caption{(Color online) The transmission probability  $T$ across
the superlattice as a function of the incoming electron energy for: (a) $N=40$; (b)$N=100$.
The following paramenters
are used:
$\phi=3\pi/4$, the ripple radius $R=18$\AA{}, the distance between the neighbouring ripples $L_1=65$\AA{}.  }
\label{fig3}
\end{figure}

For the sake of illustration, we assume that electrons move from the left
to the right side of the superlattice. We observe  that the transmission of electrons with the {\it spin down}
is suppressed  strongly in contrast to the one for the electrons with the {\it spin up} at the energy interval
$0<E<0.2$ eV. The larger the number of ripples (with the same radius, angle, and
the distance between ripples), the stronger the suppressing effect [compare Figs.\ref{fig3}(a),(b)].
Thus, the number of ripples enables to us to control the suppression degree of
the electron transmission with the spin down in the energy interval $0<E<0.2$ eV.

Note, that the increasing number of ripples in the superlattice gives rise to the periodic
suppression of the transmission with the {\it spin up} as well.
This result can be understood as the quantum interference of the two phase factors,
 that are brought about by the ripple ($e^{im\phi}$) and the flat graphene piece ($e^{ikL_1}$).
 Indeed, the wave, incoming from the left boundary of the superlattice,
 defines a transmission coefficient $t_l(N^\prime) e^{i(kL_1+m\phi)}$ that depends on the number $(N^\prime)$
 of the elementary units from the left boundary of the superlattice with N units $(N>N^\prime)$.
 On the other hand, we have the additional contribution that is defined by the reflected
 wave (the reflected electron flow from the amount of ripples located on the right side of the superlattice)
 from the right boundary, and it has the form $t_r(N^\prime+1) e^{-i(kL_1+m\phi)}$
 The total transmission probability
from the left to the right side is determined by the formula
$|t|^2=|t_l(N^\prime) e^{i(kL_1+m\phi)}+t_r(N^\prime+1)
e^{-i(kL_1+m\phi)}|^2= |t_l(N^\prime)
e^{i2(kL_1+m\phi)}+t_r(N^\prime+1)|^{2}$. The minimum of the
probability can be expressed in the form
\begin {equation}
2[m_{\pm}(E)\phi +k L_{1}]=\pi (2n+1)\,, \quad n=0,1,2,\dots
\end{equation}
By means of Eqs.(\ref{5},\ref{12}), we arrive to the condition
\begin {equation}
\left(\frac{R\phi}{\gamma}\sqrt{(E-s\lambda_{x})^{2}-\lambda_{y}^{2}}
+\frac{E}{\gamma} L_{1}\right)=\pi (n+1/2)\ ,\ s=\pm\,,
\end{equation}
which yields
\begin{equation}
\label{24}
E\left(R\phi\sqrt{(1-s\lambda_{x}/E)^{2}-(\lambda_{y}/E)^{2}} +
L_{1}\right)=\gamma\pi (n+1/2)\,.
\end{equation}

At the considered energies the following condition holds:
$E\gg \lambda_{x},\lambda_{y}$. It leads us immediately to the quantization
condition for the energy, when the suppression takes place:
\begin {equation}
E_{n}\approx\frac{\gamma\pi}{R\phi +L_{1}}(n+1/2)\,, \quad n=0,1, 2, \dots
\end{equation}
Consequently, we find that the suppression occurs with the energy step
\begin {equation}
\label{period}
\Delta E=E_{n+1}-E_{n}=\frac{\gamma\pi}{R\phi +L_{1}}
\end{equation}
as a function of the ripple radius $R$, the angle $\phi$, and the distance $L_1$.

This simple analysis confirms that the numerical results for the transmission minima
do not depend on the electron spin polarization at
large enough energies  $E$ of the incoming electrons.
On the other hand, when the energy $E\sim \lambda_{x}$,
the term $s\lambda_{x}/E$  [see Eq.(\ref{24})]  affects the transmission.
In this case, there is a strong dependance of the transmission coefficient on the spin
polarization as is demonstrated by the numerical computations.
\begin{figure}
\includegraphics[height=6cm,clip=]{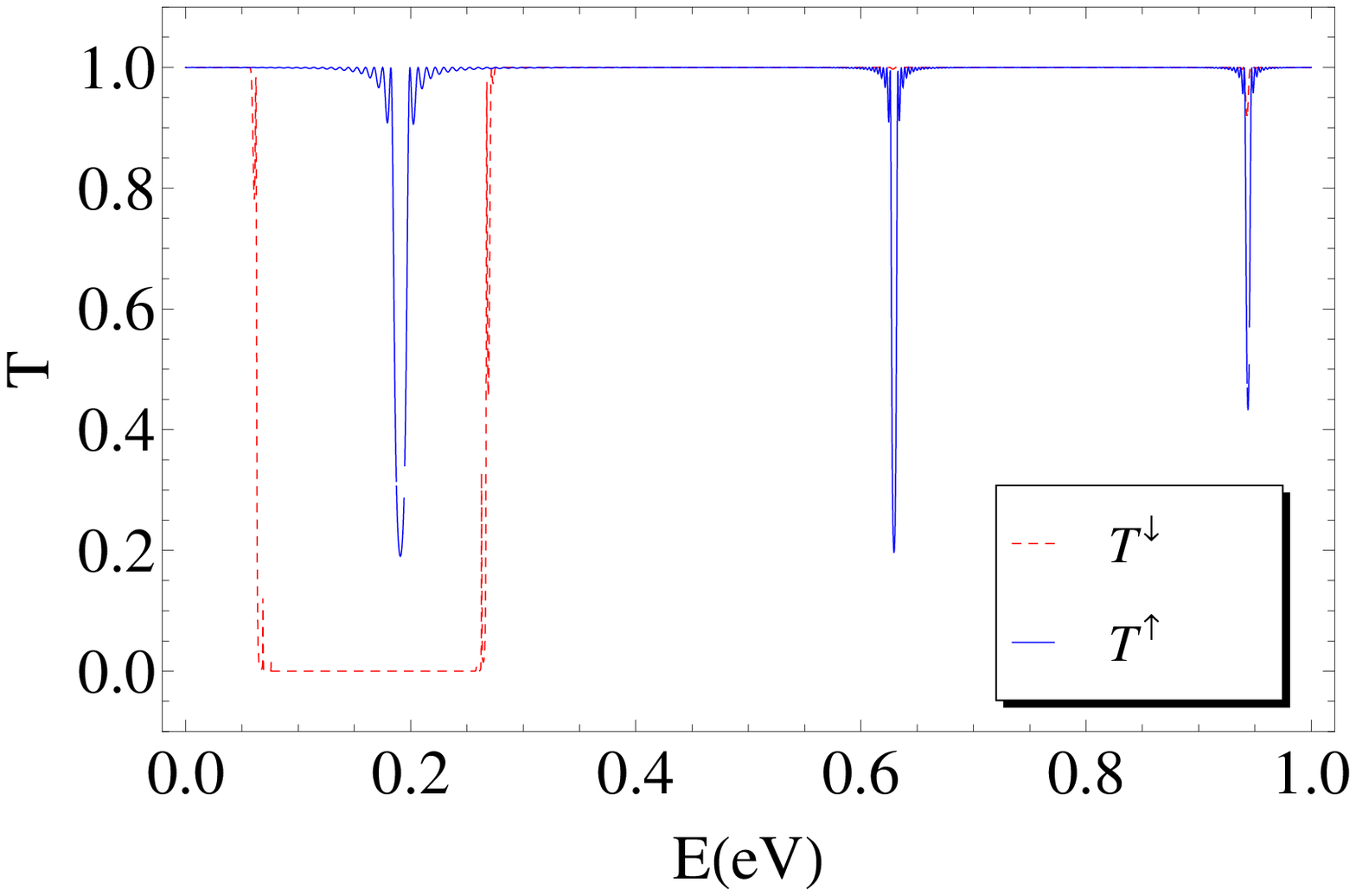}
\begin{picture}(0,0)(0,0)
\put(-65,130){\makebox(0,0)[b]{$(a)$}}
\end{picture}\
\includegraphics[height=6cm,clip=]{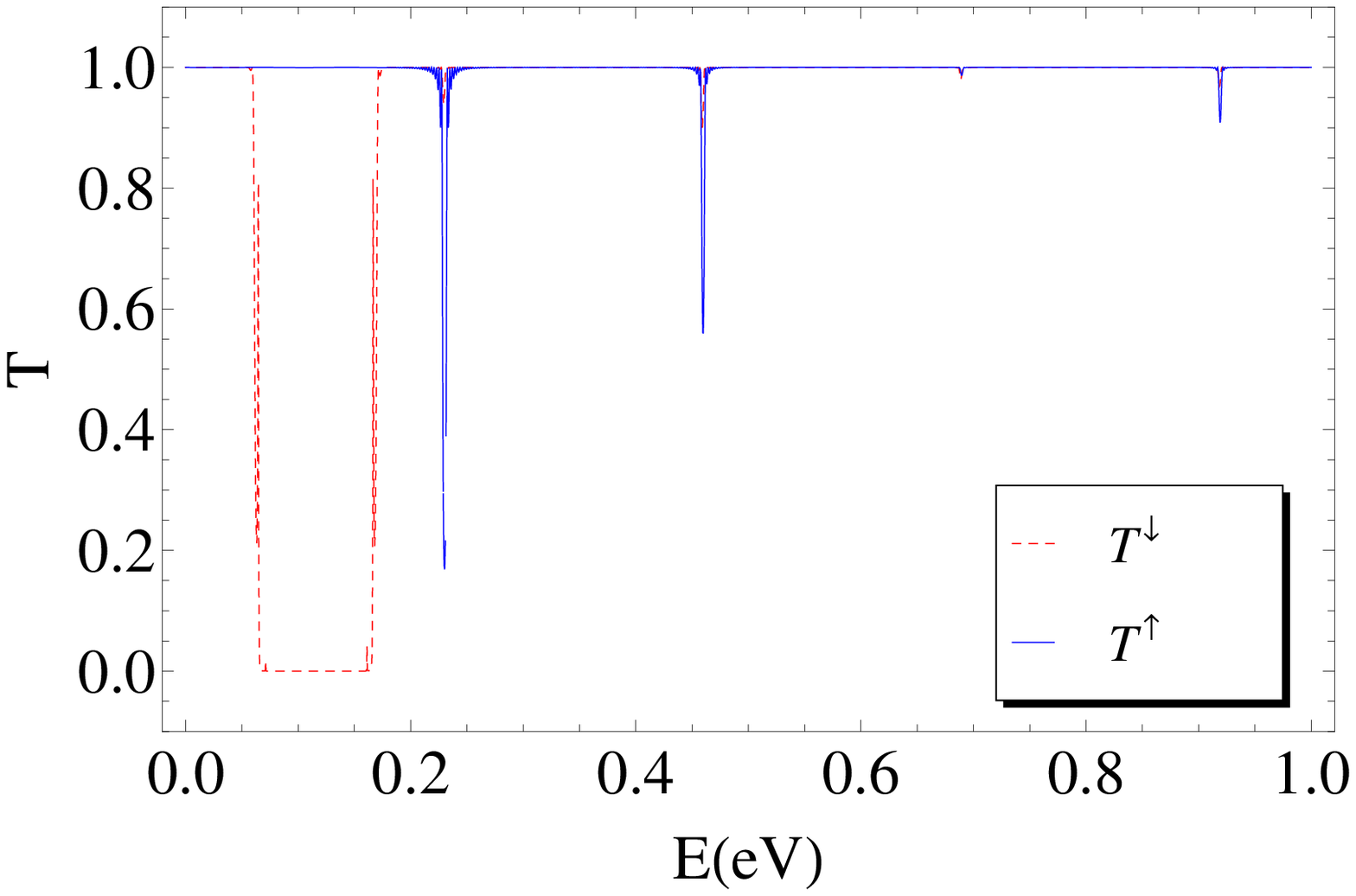}
\begin{picture}(0,0)(0,0)
\put(-65,130){\makebox(0,0)[b]{$(b)$}}
\end{picture}\\
\caption{(Color online)
The transmission probability  $T$ across
the superlattice as a function of the incoming electron energy for: (a) $R=8$ \AA{}; (b)$R=18$ \AA{}.
The following parameters
are used: the number of ripples $N=100$,
$\phi=3\pi/4$, the distance between the neighbouring ripples $L_1=45$\AA{}.
}
\label{fig4}
\end{figure}

Keeping in mind, that all observed effects exhibit themselves at
large number of ripples, we observe similar pattern (compare
Fig.\ref{fig3}, \ref{fig4}) for the transmission as a function of
the incoming electron energy and the ripple radius. Note, however,
that the decrease of the ripple radius increases the suppression of
the transmission of the electrons with the {\it spin down}. The
smaller the ripple radius, the larger the filtering effect in the
energy interval $0.05<E<0.3$ eV. This fact is due to the inverse
dependence of the parameters $\lambda_{x}$ and $ \lambda_{y}$ on the
radius [see Eq.(\ref{def})], that are responsible for the spin-orbit
effect in the corrugated structure. The observed periodicity
confirms remarkable well the validity of Eq.(\ref{period}), that
describes the suppression effect for the ballistic electrons.

\begin{figure}
\includegraphics[height=6cm,clip=]{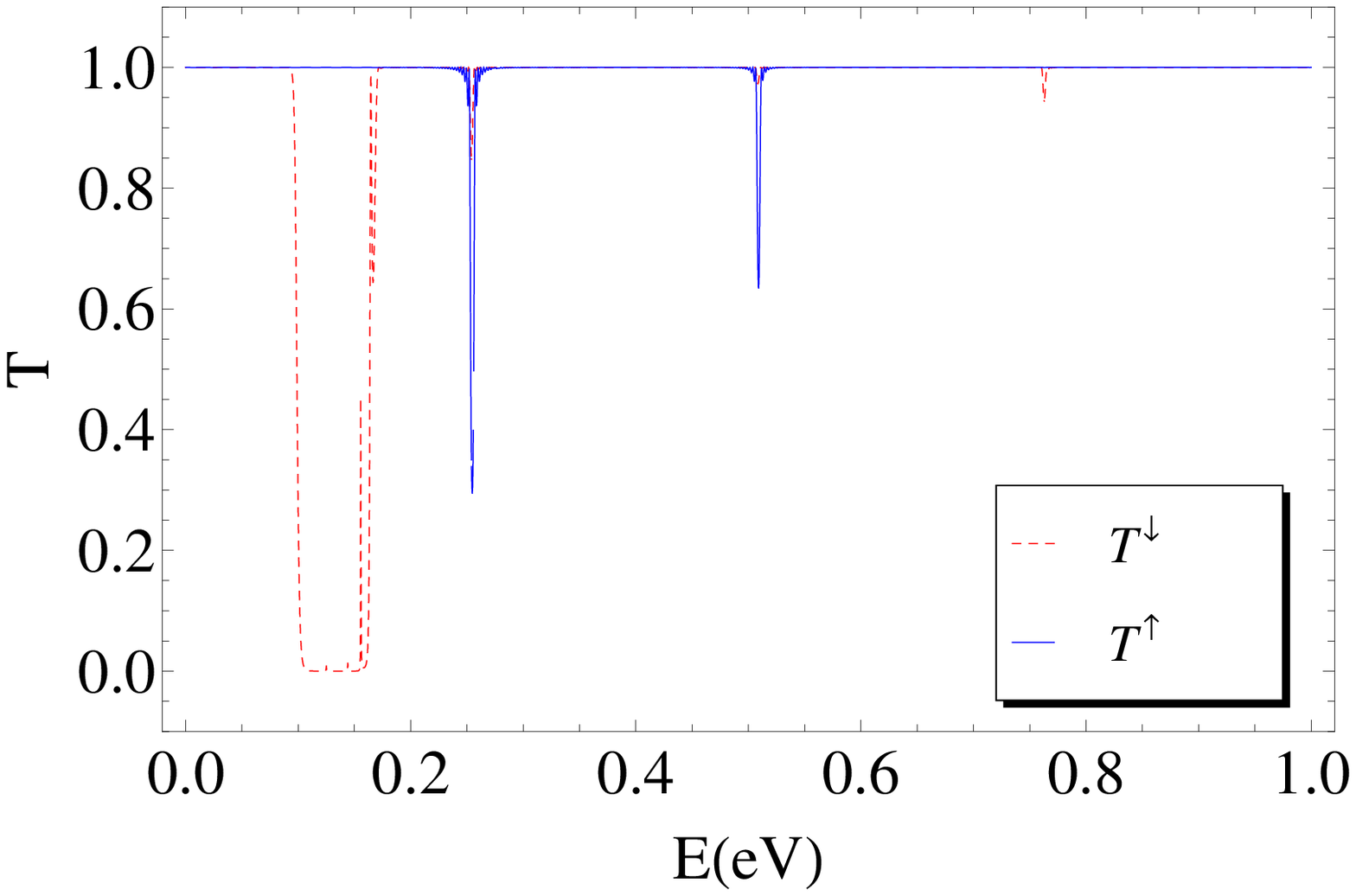}
\begin{picture}(0,0)(0,0)
\put(-65,130){\makebox(0,0)[b]{$(a)$}}
\end{picture}\
\includegraphics[height=6cm,clip=]{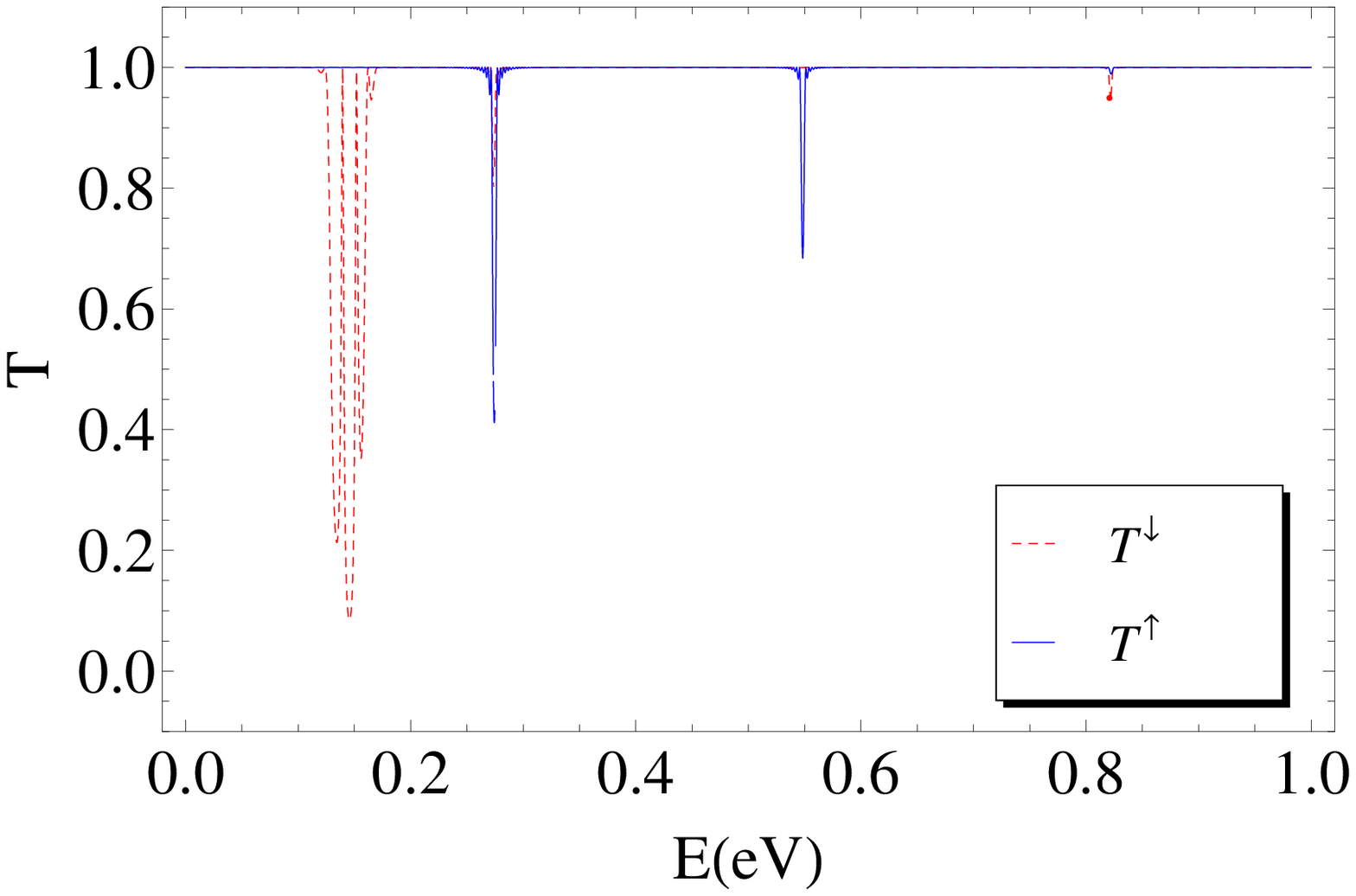}
\begin{picture}(0,0)(0,0)
\put(-45,130){\makebox(0,0)[b]{$(b)$}}
\end{picture}\
\caption{(Color online)
The transmission probability  $T$ across
the superlattice as a function of the incoming electron energy for: (a)$\phi =3\pi/5$; (b)$\phi =\pi/2$.
The following parameters
are used: the number of ripples $N=100$; the ripple radius $R=18$ \AA{};
 the distance between the neighbouring ripples $L_1=45$\AA{}.
}
\label{fig5}
\end{figure}

Another variable, that affects the transmission phenomenon, is the
ripple angle. It seems that the decrease of the ripple angle affects
strongly the transmission phenomenon discussed above. With the
increase of the ripple angle at the fixed radius (compare
Figs.\ref{fig4}b,\ref{fig5}) the filtering effect becomes more
pronounced. It appears that the smoother ripple curvature (the
larger is the angle, the larger the arc length $L_2\approx R\phi$),
the larger the filtering effects. Again, Eq.(\ref{period}) describes
remarkably well the periodicity of the suppressing.

\begin{figure}
\includegraphics[height=6cm,clip=]{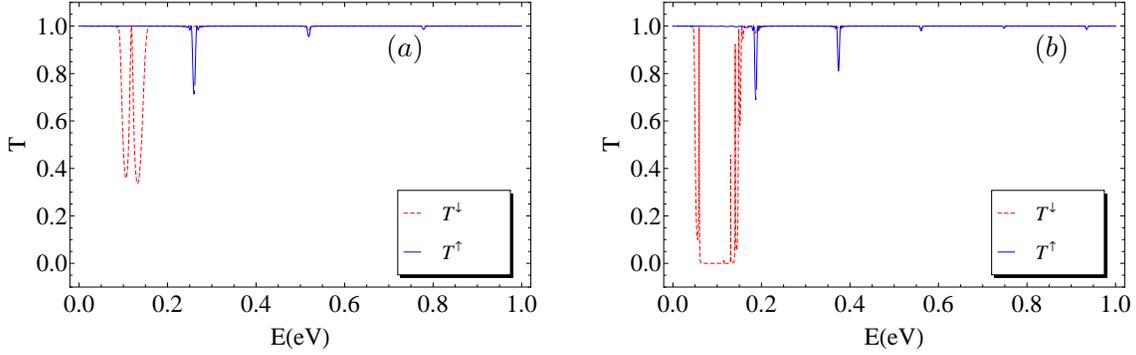}
\begin{picture}(0,0)(0,0)
\put(-65,130){\makebox(0,0)[b]{$(a)$}}
\end{picture}\
\includegraphics[height=6cm,clip=]{rip18x65N40Phi3pi4.eps}
\begin{picture}(0,0)(0,0)
\put(-45,130){\makebox(0,0)[b]{$(b)$}}
\end{picture}
\caption{(Color online)
The transmission probability  $T$ across
the superlattice as a function of the incoming electron energy for
 the distance between the neighbouring ripples:
(a)$L_{1}=35$ \AA{}; (b)$L_{1}=65$\AA{}. The following parameters
are used: the number of ripples $N=40$; $\phi=3\pi/4$; the ripple
radius $R=18$ \AA{}.}
 \label{fig6}
\end{figure}

And finally, the effect of the distance between the neighbouring
ripples yields the filtering effect in the energy interval
$0.05<E<0.15$ eV (see Fiq.\ref{fig6}). It seems the sparse rippled systems suppress the
filtering effect. Indeed, in this case there is a tendency to the
limit of the flat graphene where the filtering effect does not exist
at all.

\section{Summary}

The coupling between the geometry and electron propagation in graphene
sheet provides novel interesting phenomena.  The nonzero curvature of the
corrugated graphene region enhances the relatively weak spin-orbit
interaction of the carbon atom. As a result of this interplay, we found
the spin dependent electron transmission through the corrugated graphene system.
Two cases of the ripple arrangement
are investigating. In the first case, the
interference between successive scattering events is neglecting,
i.e., scattering events are treated as independent. This case can be
interpreted as a natural graphene system with the disorder in the
ripples positions. The transport of electrons across such a system is
characterizing by the mean free path, that, in our case, is the spin
dependent. In the second case, it is assumed that there is a
possibility to create a periodically repeated structure of the
rippled pattern in the graphene sheet. There exist the energy
regions, transparent for electrons with the one spin polarization, while
electrons with the opposite spin polarization are fully reflected.
The situation becomes inverse, once we change the flow direction
through our superlattice. The effect can be enhanced with the increasing of number
elements in our superlattice. On the other hand, we can control
the energy windows by altering the ripple characteristics: the radius, the angle,
 the length of the flat region between ripples. We found that the smaller the ripple radius,
 the larger the filtering effect at $N>>1$. The smoothness of the ripple can also
 support this effect.
In our opinion, the corrugations are most possible sources of
scattering phenomenon in graphene. The presence of the ripples can be the major
reason that explains the finite the mean free path of the electrons in graphene,
which can depend of the spin polarization. And corrugations, that can be controlled
externally, could open a new avenue in nanoelectronics based on graphene based
physics.

\section*{Acknowledgments}
This work is supported by the Slovak Academy of Sciences in the
framework of VEGA Grant No. 2/0009/19.

\end{document}